\documentclass[prd, preprint, nofootinbib]{revtex4-1}
\usepackage[dvipdfmx]{graphicx}
\usepackage{amsfonts}
\usepackage{latexsym}
\usepackage{amssymb}
\usepackage{epsfig}
\usepackage{comment}
\usepackage{here}	
\usepackage{amsmath}

\def\slashb#1{\not\!\!#1}
\newcommand{\im}[1]{\text{Im}\,#1}

\begin{document}

\title{Modular symmetry anomaly  \\
in magnetic flux compactification}

\author{Yuki Kariyazono, Tatsuo Kobayashi, Shintaro~Takada, Shio Tamba, and Hikaru Uchida}
 \affiliation{
Department of Physics, Hokkaido University, Sapporo 060-0810, Japan}



\begin{abstract}
We study modular symmetry anomalies in four-dimensional low-energy effective field theory, which is derived from six-dimensional 
supersymmetric $U(N)$ Yang-Mills theory by magnetic flux compactification.
The gauge symmetry $U(N)$ is broken to $U(N_a) \times U(N_b)$ 
by magnetic fluxes.
It is found that 
Abelian subgroup of the modular symmetry corresponding to discrete part of $U(1)$ can be anomalous, 
but other elements independent of $U(1)$ in the modular symmetry are always anomaly-free.  
\end{abstract}

\preprint{EPHOU-19-007}
\preprint{}

\vspace*{3cm}
\maketitle



\section{Introduction}

The modular symmetry is a geometrical feature, which torus compactification as well as 
orbifold compactification has.
Furthermore, the modular symmetry plays an important role in 
four-dimensional (4D) low-energy effective field theory derived from higher dimensional field theory 
and superstring theory.

The modular symmetry in string-derived supergravity theory was studied 
in Ref.~\cite{Ferrara:1989bc} and also its anomaly was studied 
in  Ref.~\cite{Derendinger:1991hq,Ibanez:1992hc}. (See also for anomalies in explicit heterotic 
orbifold models Ref.~\cite{Araki:2007ss}.)
Recently, these studies were extended to supergravity theory derived by 
magnetized and intersecting D-brane models \cite{Kobayashi:2016ovu}.
Furthermore, their anomalies are also interesting from the phenomenological viewpoint \cite{Ibanez:1992hc,Ibanez:1991zv,Kawabe:1994mj}.

Also it was studied how massless modes transform under modular symmetry in heterotic orbifold models
 \cite{Lauer:1989ax,Lerche:1989cs,Ferrara:1989qb}.
 Recently, modular transformation behavior of massless modes was studied in magnetized D-brane models 
 as well as intersecting D-brane models \cite{Cremades:2004wa,Kobayashi:2017dyu,Kobayashi:2018rad,Kobayashi:2018bff}.
 Then, it was found that the modular symmetry transforms massless modes each other, and 
 that is a sort of flavor symmetries.
 On the other hand, it was shown that non-Abelian discrete flavor symmetries appear in heterotic orbifold models 
\cite{Kobayashi:2006wq,Kobayashi:2004ya,Ko:2007dz,Beye:2014nxa,Olguin-Trejo:2018wpw,Nilles:2018wex}
and magnetized/intersecting D-brane models 
\cite{Abe:2009vi,Abe:2009uz,BerasaluceGonzalez:2012vb,Marchesano:2013ega,Abe:2014nla,Higaki:2005ie} 
through analysis independent of modular symmetry.
 Indeed, a relation between modular symmetry and non-Abelian discrete flavor symmetry was also 
 studied \cite{Kobayashi:2018rad}.(See also Ref.~\cite{Baur:2019kwi}.)

Non-Abelian discrete flavor symmetries are interesting from the phenomenological viewpoints 
\cite{Altarelli:2010gt,Ishimori:2010au,King:2013eh}.
Various finite groups have been utilized such as 
$S_3, A_4, S_4, A_5$, etc. for  4D field-theoretical model building.
Then, many models have been proposed 
in order to realize quark and lepton masses and their mixing angles and CP phases.
The modular group includes  $S_3, A_4, S_4, A_5$ as its finite subgroups \cite{deAdelhartToorop:2011re}.
This aspect in addition to the above string compactification inspired a new approach of 
4D field-theoretical model building \cite{Feruglio:2017spp}, where 
finite subgroups of the modular symmetry are used as non-Abelian discrete flavor symmetries and 
also couplings and masses are assumed to transform non-trivially under such finite subgroups.
Such a new approach has been applied to models with $S_3, A_4, S_4, A_5$  modular symmetries 
\cite{Kobayashi:2018vbk,Penedo:2018nmg,Criado:2018thu,Kobayashi:2018scp,Novichkov:2018ovf,Okada:2018yrn,Kobayashi:2018wkl,Novichkov:2018yse,Ding-King-Liu}.

Thus, the modular symmetry is important from both theoretical and phenomenological viewpoints.
In general, continuous and discrete symmetries can be anomalous.
(See for anomalies of Abelian and non-Abelian discrete symmetries 
Refs.~\cite{Krauss:1988zc,Ibanez:1991hv,Banks:1991xj,Araki:2008ek}.)
Anomalous symmetries can be broken by non-perturbative effects.
That is, breaking terms are induced by non-perturbative effects.
Such breaking terms may have important implications.
The purpose of this paper is to study the anomaly structure of the modular symmetry 
in 4D low-energy effective field theory derived from 
magnetic flux compactification of higher dimensional supersymmetric Yang-Mills theory, 
which is effective field theory of magnetized D-brane models.

This paper is organized as follows.
In Sec.~\ref{sec:magne-D}, we present our setup and give a brief review 
on magnetic flux compactification and the modular transformation of zero-modes.
In Sec.~\ref{sec:anomaly}, we study the anomaly structure of the modular symmetry.
Sec.~\ref{sec:conclusion} is our conclusion.

\section{Modular transformation of magnetic flux compactification}
\label{sec:magne-D}

\subsection{Setup and wavefunctions}

Here, we present our setup and give a brief review on magnetic flux compactification.
We start with six-dimensional supersymmetric $U(N)$  Yang-Mills theory, which can be derived from D-brane system.
Then, we consider the two-dimensional torus $T^2$ compactification with magnetic flux. 
Similarly, we can study higher dimensional theory such as ten-dimensional supersymmetric Yang-Mills theory on 
$T^2\times T^2 \times T^2$, which can also be derived from D-brane system.
Indeed, magnetic flux compactification on $T^2$ leads to tachyonic modes.
On the other hand, tachyonic modes can be avoided on $T^2\times T^2 \times T^2$ by 
choosing proper combinations of magnetic fluxes and sizes of three $T^2$ tori.
However, each $T^2$ tours  is important to study the anomaly structure.
Thus, here we  concentrate on the $T^2$ compactification by assuming absence of tachyonic modes 
in $T^2\times T^2 \times T^2$.

We use the complex coordinate $z= x^1+ \tau x^2$ on $T^2$, 
where 
$\tau$ is the complex modulus parameter, and $x^1$ and $x^2$ are real coordinates.
The metric on $T^2$ is given by 
\begin{equation}
g_{\alpha \beta} = \left(
\begin{array}{cc}
g_{zz} & g_{z \bar{z}} \\
g_{\bar{z} z} & g_{\bar{z} \bar{z}}
\end{array}
\right) = (2\pi R)^2 \left(
\begin{array}{cc}
0 & \frac{1}{2} \\
\frac{1}{2} & 0
\end{array}
\right) .
\end{equation}
We identify $z \sim z +1$ and $z \sim z + \tau$ on $T^2$.

We introduce the following magnetic flux along the diagonal direction,
\begin{equation}
F = i\frac{\pi}{\im{\tau}}  (dz \wedge d\bar{z})
\left(
\begin{array}{cc}
M_a\mathbb{I}_{N_a \times N_a} & 0 \\
0 &  M_b\mathbb{I}_{N_b \times N_b}
\end{array} \right),
\end{equation}
where $N_a+N_b=N$, $\mathbb{I}_{N_{a,b} \times N_{a,b}}$ denotes the $(N_{a,b} \times N_{a,b})$ identity matrix 
and $M_{a,b}$ must be integer.
This form of magnetic flux corresponds to the  vector potential,
\begin{equation}
\label{eq:A}
A(z) = \frac{\pi }{\im{\tau}} \im{(\bar{z}dz)}
\left(
\begin{array}{cc}
M_a\mathbb{I}_{N_a \times N_a} & 0 \\
0 &  M_b\mathbb{I}_{N_b \times N_b}
\end{array} \right).
\end{equation}
Because of this gauge background, the $U(N)$ gauge symmetry breaks to $U(N_a) \times U(N_b)$.

Now let us study the gaugino sector.
The spinor field on $T^2$  has two components, $\lambda_\pm$.
They are decomposed to 
\begin{equation}\left(
\begin{array}{cc}
\lambda^{aa}_\pm & \lambda^{ab}_\pm \\
\lambda^{ba}_\pm & \lambda^{bb}_\pm
\end{array}\right).
\end{equation}
Here $\lambda^{aa}$ and $\lambda^{bb}$ correspond to 
the gaugino fields of unbroken gauge groups, $U(N_a)$ and $U(N_b)$, respectively, 
while $\lambda^{ab}$ and $\lambda^{ba}$ correspond to $(N_a,\bar N_b)$ and 
 $(\bar N_a, N_b)$ under $U(N_a) \times U(N_b)$.

The zero-mode equation with the above gauge background (\ref{eq:A}),
\begin{equation}
i \slashb{D} \lambda_{\pm} = 0,
\end{equation} 
has chiral solutions.
When $M=M_a-M_b$ is positive, $\lambda^{ab}_{+}$ and $\lambda^{ba}_-$ have 
$M$ degenerate zero-modes,\footnote{
Note that the six-dimensional chirality is fixed.
Then, $\lambda^{ab}_{+}$ and $\lambda^{ba}_-$ are combined with 4D left-handed and 
right-handed spinor fields, and they correspond to a pair of matter and anti-matter.}
 whose profiles are written by \cite{Cremades:2004wa}
\begin{equation}
\psi^{j,M}_{T^2}(z) = \mathcal{N} e^{i\pi  M z \frac{\im{z}}{\im{\tau}}} \cdot \vartheta \left[
\begin{array}{c}
\frac{j}{M} \\
0
\end{array}
\right] \left( M z, M\tau \right),
\end{equation}
with $j=0,1,\cdots, (M-1)$, 
where $\vartheta$ denotes the Jacobi theta function, 
\begin{equation}
\vartheta \left[
\begin{array}{c}
a \\
b
\end{array}
\right] (\nu, \tau) = \sum_{l \in {\bf Z}} e^{\pi i (a+l)^2 \tau} e^{2 \pi i (a+l)(\nu+b)} .
\end{equation}
Here, $\mathcal{N}$ denotes  the normalization factor given by 
\begin{equation}
\label{eq:normalization}
\mathcal{N} = \left( \frac{2\im{\tau} M}{\mathcal{A}^2} \right)^{1/4}, 
\end{equation}
with $\mathcal{A}= 4 \pi^2 R^2 \im{\tau}$.

On the other hand, when $M$ is negative, $\lambda^{ab}_{-}$ and $\lambda^{ba}_+$ have 
$|M|$ degenerate zero-modes, whose profiles  are the same as $\psi^{j,M}(z)$ 
except $M$ replacing by $|M|$.
Hereafter, we set $M$ to be positive.
That is, we consider the model that has $M$ degenerate zero-modes of 
$\lambda^{ab}_{+}$ and $\lambda^{ba}_-$, but no zero-modes of $\lambda^{ab}_{-}$ and $\lambda^{ba}_+$.

Because of the chiral spectrum, $U(1)_a$ and $U(1)_b$ are anomalous 
in 4D low-energy effective field theory.
For example, both the mixed anomalies, $U(1)_a-SU(N_b)^2$ and $U(1)_b-SU(N_a)^2$ are 
proportional to $M$.
Such anomalies can be canceled by the Green-Schwarz mechanism, if we include the 
Green-Schwarz field in our theory.
The Green-Schwarz mechanism cancels anomalies by the shift of axions$\chi_{a,b}$, 
\begin{equation}\label{eq:axion-shift}
\chi_{a,b} \rightarrow \chi_{a,b} + \alpha_{a,b},
\end{equation}
under $U(1)_{a,b}$ transformation, where $\alpha_{a,b}$ are $U(1)_{a,b}$ gauge transformation 
parameters  \cite{Aldazabal:2000dg}.
Those axions are eaten by $U(1)_{a,b}$ gauge bosons and then $U(1)_{a,b}$ gauge bosons become massive.

Similarly, this theory has the $U(1)-({\rm gravity})^2$ and $U(1)^3$ anomalies.
Those can also be canceled by the Green Schwarz mechanism.

In the next section, we will study the $T^2/Z_2$ orbifold background.
For simplicity, we focus on orbifold models without Wilson lines.
The zero-mode wavefunctions on $T^2/Z_2$ are obtained from the above wavefunctions \cite{Abe:2008fi}.
The above wavefunctions have the following property:
\begin{equation}
\psi^{j,M}_{T^2}(-z) = \psi^{M-j,M}_{T^2}(z).
\end{equation}
Thus, the $T^2$ wavefunction with $j=0$ is still the $Z_2$-even zero-mode on $T^2/Z_2$.
Also, when $M=$ even, the $T^2$ wavefunction with $j=M/2$ is still the $Z_2$-even zero-mode on $T^2/Z_2$.
That is, we obtain 
\begin{equation}
\psi^{j,M}_{T^2/Z^+_2}(z) = \psi^{j,M}_{T^2}(z),
\end{equation}
for $j=0,M/2$.
For the other, the $Z_2$-even and odd zero-modes can be written by 
 \begin{equation}
\psi^{j,M}_{T^2/Z^\pm_2}(z) = \frac{1}{\sqrt 2}\left( \psi^{j,M}_{T^2}(z) \pm \psi^{M-j,M}_{T^2}(z) \right). 
\end{equation}
When $M=$ even, totally the numbers of $Z_2$-even and odd zero-modes are equal to $(M/2+1)$ 
and  $(M/2-1)$, respectively.
When $M=$ odd, the   numbers of $Z_2$-even and odd zero-modes are equal to $((M-1)/2+1)$ 
and  $((M-1)/2)$, respectively.
Either $Z_2$ even or odd modes are projected out by the $Z_2$ projection.

The anomalies of $U(1)_a$ and $U(1)_b$ on the $T^2/Z_2$ orbifold, e.g. for the 
$Z_2$-even modes $\psi^{j,M}_{T^2/Z^+_2}(z)$,  can be studied 
in the same way as on the torus.
Those anomalies can also be canceled by the Green-Schwarz mechanism.

\subsection{Modular transformation}

Here, we give a brief review on modular transformation of zero-mode wavefunctions 
\cite{Cremades:2004wa,Kobayashi:2017dyu,Kobayashi:2018rad,Kobayashi:2018bff}.
Following \cite{Kobayashi:2017dyu}, we restrict ourselves to even magnetic fluxes $M$.

Under the modular transformation, the modulus $\tau$ transforms as 
\begin{equation}
\tau \longrightarrow  \frac{a\tau + b}{c \tau + d}.
\end{equation}
This group includes two important generators, $S$ and $T$,
\begin{eqnarray}
& &S:\tau \longrightarrow -\frac{1}{\tau}, \\
& &T:\tau \longrightarrow \tau + 1.
\end{eqnarray}

The generator $S$ transforms the zero-mode wavefunctions as 
\begin{equation}
\label{eq:magne-S}
\psi^{j,M} \rightarrow \frac{1}{\sqrt{M}}\sum_k e^{2\pi i jk/M} \psi^{k,M}.
\end{equation}
On the other hand, the generator $T$ transforms the zero-mode wavefunctions  
\begin{equation}
\label{eq:magne-T}
\psi^{j,M} \rightarrow e^{ \pi i j^2/M} \psi^{j,M}.
\end{equation}
Generically, the $T$ generator satisfies \cite{Kobayashi:2017dyu}
\begin{equation}
T^{2M} = \mathbb{I}_{ M\times M },
\end{equation}
on the zero-modes, $\psi^{j,M}$.
Furthermore, in Ref. \cite{Kobayashi:2017dyu} it is shown that 
\begin{equation}
(ST)^3 = e^{\pi i/4}\mathbb{I}_{M\times M },
\end{equation}
on the zero-modes, $\psi^{j,M}$.
Hence, $T$ and $(ST)^3$ are represented by diagonal matrices on $\psi^{j,M}$, and 
they are $Z_{2M}$ and $Z_8$ symmetries, respectively.

The above representations of $S$ and $T$ on $\psi^{j,M}$ are reducible.
It is obvious that $\psi^{j,M}$ and $\psi^{M-j,M}$ transform in the same way 
under both $S$ and $T$.
That implies that the orbifold basis $\psi^{j,M}_{T^2/Z^\pm_2}(z)$ corresponds to 
 the irreducible representation.
We denote such irreducible representations by $S_\pm$ and $T_\pm$.
Their explicit forms can be read off from the above representations of $S$ and $T$.
Note that when $M=$ even, $S_+$ and $T_+$ are $(M/2+1)\times (M/2+1)$ matrices, and 
 $S_-$ and $T_-$ are $(M/2-1)\times (M/2-1)$ matrices.

\section{Modular symmetry anomaly}
\label{sec:anomaly}

Here, we study the modular symmetry anomaly.
Anomalies of non-Abelian discrete symmetries were studied in Ref.~\cite{Araki:2008ek}.
Each element of a non-Abelian discrete group, $g$, generates Abelian discrete symmetry,  $Z_K$ 
i.e. $g^K=1$.
Thus, basically anomalies of non-Abelian discrete group are studied by 
analyzing Abelian discrete anomalies of each element, $g$.
However, states correspond to a multiplet under a non-Abelian discrete symmetry.
That is, $g$ is represented by a matrix.
Suppose that zero-modes correspond to the (anti-)fundamental representation of $SU(N_b)$.
Then, if $\det g =1$, the mixed $Z_K-SU(N_b)^2$ anomaly vanishes.
Otherwise, the $Z_K$ symmetry generated by $g$ can be anomalous.
Furthermore, suppose that zero-modes correspond to the bi-fundamental representation 
$(N_a,\bar N_b)$ under $SU(N_a)\times SU(N_b)$.
Then, if $\det g^{N_a} =1$, the mixed $Z_K-SU(N_b)^2$ anomaly vanishes.
Otherwise, the $Z_K$ symmetry generated by $g$ is anomalous.
Hence, the quantity $\det g$ is important to examine anomalies.
If  $\det g \neq 1$, such 
discrete symmetry can be anomalous.
Also, we can study $Z_K-({\rm gravity})^2$ anomalies.
If $\det g=1$, such elements do not contribute to gravitational mixed anomalies.

\subsection{$T^2/Z_2$ orbifold}

As mentioned above, the orbifold basis is more fundamental.
Thus, we first study anomalies due to the $Z_2$-even modes on the $T^2/Z_2$ orbifold.
Here, we study anomalies by examining $\det g$  for smaller $M$ concretely.

\subsubsection{$M=2$}

Here, we study the modular symmetry for $M=2$.
Note that the zero-modes on $T^2$ are the same as the $Z_2$-even zero-modes 
on $T^2/Z_2$.
First, we study diagonal elements, $T$ and $(ST)^3$.
Their explicit forms are written as 
\begin{equation}
T_{(2)}=\left(
\begin{array}{cc}
1 &  \\
 & i
\end{array}\right), \qquad 
(S_{(2)}T_{(2)})^3 = e^{\pi i /4}\mathbb{I}_{2 \times 2},
\end{equation}
where we have omitted vanishing off-diagonal entries.
That is the $Z_4 \times Z_8$ symmetry, 
and they satisfy $\det T_{(2)} \neq 1$ and $\det (S_{(2)}T_{(2)})^3 \neq 1$.
Thus, both symmetries can be anomalous.
However, their combination, 
\begin{equation}
T'_{(2)}=T_{(2)}(S_{(2)}T_{(2)})^{-3}=\left(
\begin{array}{cc}
e^{-\pi i /4} &  \\
 & e^{\pi i /4}
\end{array}
\right),
\end{equation}
has $\det T'_{(2)}=1$ and is always anomaly-free.
This is the $Z_8$ symmetry.
Hence, the  $Z_4 \times Z_8$ symmetry can be broken to 
$Z_8$ by anomalies.
The generator $A_{(2)}=(S_{(2)}T_{(2)})^3 $ can be anomalous.
Note that $(A_{(2)})^4=(T'_{(2)})^4$.
It is obvious that $A_{(2)}$ is commutable with any element.
Therefore, at least the elements $(A_{(2)})^kg$~~($k=1,2,3$) with $\det g =1$ 
has  $\det((A_{(2)})^kg) \neq 1$ and can be anomalous  
among all of the elements, which are generated by $S_{(2)}$ and $T_{(2)}$.
Indeed, explicit calculation shows that 
the order of the full group generated by $S_{(2)}$ and $T_{(2)}$ is equal to 
192, and among them the number of elements with $\det g =1$ is equal to 48.
Thus, all of the elements with $\det h \neq 1$ can be written by 
$h=(A_{(2)})^kg$~~($k=1,2,3$) with $\det g =1$.
That is, only the element  $A_{(2)}$ is important for anomalies.

The element $A_{(2)}$ can be anomalous.
For example, it can lead to the mixing anomalies with $SU(N_a)$ and $SU(N_b)$.
However, it is remarkable that the  element $A_{(2)}$ corresponds to 
a subgroup of $U(1)_a$ as well as $U(1)_b$.
Thus, when we include the Green-Schwarz field in our theory in order to cancel $U(1)$ anomalies,  
the discrete anomalies corresponding to $A_{(2)}$ can also be canceled 
by the same Green-Schwarz mechanism as one for $U(1)_a$ and $U(1)_b$.
The other discrete parts, which are independent of $A_{(2)}$, are always anomaly-free.

Similarly, we can study the $Z_K-({\rm gravity})^2$.
Only the element $A_{(2)}$ can lead to such gravitational mixed anomalies, 
because the others have $\det g =1$.
Such anomalies can be canceled by the  same Green-Schwarz mechanism as one for $U(1)_a$ and $U(1)_b$.

The $(Z_K)^3$ anomaly has a clear meaning only if $Z_K$ originates from $U(1)$ group \cite{Krauss:1988zc,Ibanez:1991hv,Banks:1991xj,Araki:2008ek}.\footnote{When we examine anomalies by the Feynman diagram calculations, 
we use currents associated with symmetries, but we can not define currents for discrete symmetries.
On the other hand, we can examine anomalies of discrete symmetries by the path integral approach.
Then, anomalies appear as mixed anomalies between a discrete symmetry and gauge symmetries (gravity), whose gauge bosons (gravitons) 
are included in covariant derivatives of fermions.}
Thus, the relation between  $(Z_K)^3$ and $U(1)^3$ anomalies as well as their Green-Schwarz cancellation mechanisms 
is rather clear when $Z_K$ is the subgroup of $U(1)$.
For the other part, we do not discuss  the $(Z_K)^3$ anomaly.

As mentioned in the previous section, in the Green-Schwarz mechanism 
the axion $\chi$ shifts under the $U(1)$ transformation to cancel anomalies.
Such an axion is the pure imaginary part of a complex field $U$ in the supersymmetric theory, 
where axionic shift (\ref{eq:axion-shift}) leads to $U \rightarrow U+i\alpha$ 
under the $U(1)$ gauge transformation with the transformation parameter $\alpha$.
It implies that $e^{-cU}$ transforms linearly and it behaves as if it has the $U(1)$ "charge" $-c$.
Non-perturbative effects such as D-brane instanton effects induce 
new terms $e^{-cU} \phi_1 \phi_2 \cdots $ in 4D low-energy effective field theory.
Such terms are invariant under the anomalous $U(1)$ and discrete symmetry with taking into account 
the transformation of $e^{-cU}$.
However, when we replace $U$ by its vacuum expectation value, such terms correspond to 
breaking terms.
Thus, breaking terms for anomalous symmetries appear.
Similar breaking terms would also appear by field-theoretical instanton effects 
even if we do not take string non-perturbative effects into account.

\subsubsection{$M=4$}

Similarly, we study the orbifold model with $M=4$, in particular the $Z_2$-even modes.
First, we study diagonal elements, $T$ and $(ST)^3$.
Their explicit forms are written as 
\begin{equation}
T_{(4)+}=\left(
\begin{array}{ccc}
1 & & \\
  & e^{\pi i/4} & \\
  & & -1
  \end{array}\right), \qquad 
(S_{(4)+}T_{(4)+})^3=  e^{\pi i /4}\mathbb{I}_{3\times 3}.
  \end{equation}
They correspond to the $Z_8 \times Z_8$ symmetry.
We find that $\det T_{(4)+} \neq 1$ and $\det (S_{(4)+}T_{(4)+})^3 \neq 1 $.
They can be anomalous.
However, their combination,
\begin{equation}
T'_{(4)+}=T_{(4)+}(S_{(4)+}T_{(4)+})^3 =\left(
\begin{array}{ccc}
e^{\pi i /4} &  &  \\
 & e^{2\pi i /4} &  \\
 &  & e^{5\pi i /4}
\end{array}
\right),
  \end{equation}
has $\det T'_{(4)+}=1$, and is always anomaly-free.
This is the $Z_8$ symmetry.
The $Z_8 \times Z_8$ symmetry can be broken to $Z_8$ by anomalies.
The generator 
$A_{(4)}=(S_{(4)+}T_{(4)+})^3  =e^{\pi i /4}\mathbb{I}_{3\times 3}$ can be anomalous again, and 
this is commutable with any element.
At least the elements $(A_{(4)+})^kg$~~($k=1,\cdots,7$) with $\det g =1$ 
has  $\det((A_{(4)+})^kg) \neq 1$ and can be anomalous  
among all of the elements, which are generated by $S_{(4)+}$ and $T_{(4)+}$.
Indeed, explicit calculation shows that the order of the full group generated by $S_{(4)+}$ and $T_{(4)+}$ is equal to 
768, and among them the number of elements with $\det g =1$ is equal to 96.
Thus, all of the elements with $\det h \neq 1$ can be written by 
$h=(A_{(4)+})^kg$~~($k=1,\cdots,7$) with $\det g =1$.

The generator $A_{(4)+}$ is a sub-element of $U(1)_a$ as well as $U(1)_b$.
Thus, anomalies originated from  $A_{(4)+}$ can be canceled by the Green-Schwarz mechanism.

\subsubsection{$M=6$}

Similarly, we study the orbifold model with $M=6$, in particular the $Z_2$-even modes.
The diagonal elements, $T$ and $(ST)^3$, are explicitly written by 
\begin{equation}
T_{(6)+}=\left(
\begin{array}{cccc}
1 & & & \\
 & e^{\pi i/6} & & \\
 & & e^{2\pi i /3} & \\
 & & & e^{3\pi i /2}
 \end{array}\right), \qquad 
 (S_{(6)+}T_{(6)+})^3 = e^{\pi i /4} \mathbb{I}_{4\times 4},
 \end{equation}
 where $\det (S_{(6)+}T_{(6)+})^3 = -1$. 
 They correspond to the $Z_{12} \times Z_{8}$ symmetry.
 They can be anomalous.
By their combinations, we can construct the diagonal elements with $\det g =1$ such as 
\begin{equation}
(S_{(6)+}T_{(6)+})^6=i\mathbb{I}_{4 \times 4}, \qquad
 (T_{(6)+})^3 (S_{(6)+}T_{(6)+})^3=\left(
\begin{array}{cccc}
e^{\pi i/4} & & & \\
 & e^{3\pi i/4} & & \\
 & &e^{\pi i/4}  & \\
 & & & e^{3\pi i/4}
 \end{array}\right),
 \end{equation}
 etc.
They include $T_{(6)+} ^k$ only for $k=3k'$ with $k'=$ integer, but 
the elements $g$ including  $T_{(6)+} ^k$  for $k=3k'+1$ and $k=3k'+2$ have $\det g \neq 1$ and can be anomalous.
The order of the above group with $\det g =1$ in the $Z_{12} \times Z_{8}$ symmetry is equal to 16.
Thus, its order reduces by the factor $1/6$.
Indeed, the order of the full group generated by $S_{(6)+}$ and $T_{(6)+}$ is equal to 
2304, and among them the number of elements with $\det g =1$ is equal to 384.
That is, the order reduces by the factor 1/6.
Here, it seems that the group elements including $T_{(6)+} ^k$ with $k=1,2$ in addition 
$(S_{(6)+}T_{(6)+})^3 $ can be anomalous.
That is different from the above cases with $M=2$ and 4.

However,  $(S_{(6)+}T_{(6)+})^3 $ corresponds to the sub-element of $U(1)_{a,b}$.
Let us combine  $T_{(6)+}$ and a discrete transformation of $U(1)_{a,b}$,
\begin{equation}
T'_{(6)+} = e^{i\alpha}T_{(6)+}.
\end{equation}
When $\alpha = -1/12$, we have $\det T'_{(6)+}=1$, and $T'_{(6)+}$ is written explicitly as 
\begin{equation}
T'_{(6)+} = \left(
\begin{array}{cccc}
e^{-\pi i /12} & & & \\
 & e^{\pi i /12} & & \\
 &  & e^{7\pi i /12} & \\
 & & & e^{-7\pi i /12} 
\end{array}\right).
\end{equation}

As a result, in the comprehensive symmetry including the modular symmetry and $U(1)_{a,b}$, 
only $U(1)_{a,b}$ including their discrete symmetries can be anomalous.
In this sense, the anomaly structure for $M=6$ is the same as the previous examples for $M=2$ and $M=4$, 
where only discrete symmetries of $U(1)_{a,b}$ as well as of course  $U(1)_{a,b}$ themselves 
can be anomalous.

\subsubsection{Larger $M$}

Similarly, we can study  anomalies for larger $M$.
The anomaly structure for larger $M$ is the same as one for $M=2,4,6$.
For $M\neq 6k$, $T_{(M)+}$ and $(S_{(M)+}T_{(M)+})^3$, in general, have $\det T_{(M)+} \neq 1$ and 
$\det  (S_{(M)+}T_{(M)+})^3 \neq 1$, although in specific values of $M$ we have 
$\det T_{(M)+} = 1$ for $(M+1)(M/2 +1) = 24k$\footnote{$M$ is obtained by $M=16n -2$ with $n$ satisfying 
$n(16n-1) = 3k$.} 
and $\det  (S_{(M)+}T_{(M)+})^3 = 1$ for $M=16k-2$.
However, we can find the element $T'_{(M)+}=T_{(M)+}(S_{(M)+}T_{(M)+})^{3m}$ satisfying 
$\det T'_{(M)+}=1$.
Then, only the element $(S_{(M)+}T_{(M)+})^3$ can be anomalous.
That is, only the discrete symmetry of $U(1)_{a,b}$ can be anomalous.

For $M=6k$, even if we combine $T_{(M)+}^\ell$ and $(S_{(M)+}T_{(M)+})^{3m}$, 
there are elements with $\det g \neq 1$ except $(S_{(M)+}T_{(M)+})^{3m}$.
However, we can obtain $T'_{(M)+} = e^{i\alpha}T_{(M)+}$ with $\det T'_{(M)+} =1$ by combining 
$ T_{(M)+} $ with a proper discrete element of $U(1)_{a,b}$.

As a result, it is found that only the  $U(1)_{a,b}$ including their discrete symmetries can 
be anomalous, but the other symmetries independent of $U(1)_{a,b}$ are always anomaly-free.

Although we have studied anomalies for the $Z_2$ even modes, 
we can study similarly anomalies for the $Z_2$ odd modes.
One example is shown in the next subsection.
Note that either $Z_2$ even or odd modes are projected out in $T^2/Z_2$ orbifold models, 
but both appear in $T^2$ models.

\subsection{$T^2$}

Similarly, we can discuss $T^2$ models.
The zero-modes of $T^2$ are combinations of $Z_2$-even and odd modes on the $T^2/Z_2$ orbifold.
For $M=2$, all of the zero-modes on $T^2$ are the $Z_2$-even zero-modes.
Thus, $S$ and $T$ are represented by $S_{(2)}$ and $T_{(2)}$.

For $M=4$, there is one $Z_2$-odd mode.
Then, the diagonal elements, $T$ and $(ST)^3$ are represented by 
\begin{equation}
T_{(4)}=\left(
\begin{array}{cc}
T_{(4)+} &   \\   & T_{(4)-} 
\end{array}\right), \qquad 
(S_{(4)}T_{(4)})^3=\left(
\begin{array}{cc}
(S_{(4)+}T_{(4)+})^3 &   \\   & (S_{(4)-}T_{(4)-})^3 
\end{array}\right),
\end{equation}
where $T_{(4)-}=e^{\pi i/4}$ and $(S_{(4)-}T_{(4)-})^3 =e^{\pi i/4}$.
That is, we have $(S_{(4)}T_{(4)})^3=e^{\pi i/4}\mathbb{I}_{4 \times 4}$.
This element corresponds to the discrete sub-group of $U(1)_{a,b}$  and can be anomalous.
Other elements independent of $U(1)_{a,b}$ discrete subgroup are always anomaly-free.
For example, from $T_{(4)}$ we can construct $T'_{(4)}=e^{i\alpha}T_{(4)}$ with $\det T'_{(4)} =1$ by 
choosing a proper value of $\alpha$.


\section{Conclusion}
\label{sec:conclusion}

We have studied the modular symmetry anomalies in magnetic flux compactifiction.
Our model is six-dimensional supersymmetric $U(N)$ Yang-Mills theory, where $U(N)$ gauge symmetry 
is broken down to $U(N_a) \times U(N_b)$ by magnetic fluxes in the compact space.
Discrete subsymmetries of $U(1)_{a,b}$ in the modular symmetry can be 
anomalous, but other discrete elements, which are independent of  $U(1)_{a,b}$, are always anomaly-free.
Anomalies of such discrete symmetries can be canceled by the same Green-Schwarz mechanism 
as the  mechanism to cancel $U(1)_{a,b}$ anomalies.
As a result, breaking terms can be induced only for continuous and discrete  $U(1)_{a,b}$ symmetries.

Here we have studied supersymmetric $U(N)$ Yang-Mills theory, which can be derived from 
D-brane models.
Similar representations of $S$ and $T$ were derived in heterotic orbifold models  \cite{Lauer:1989ax,Lerche:1989cs,Ferrara:1989qb}.
It is interesting to carry out a similar analysis on heterotic orbifold models.


\section*{Acknowledgments}
T.~K. was  supported in part by MEXT KAKENHI Grant Number JP19H04605.

%







\begin{thebibliography}{99}


\bibitem{Ferrara:1989bc} 
  S.~Ferrara, D.~Lust, A.~D.~Shapere and S.~Theisen,
  Phys.\ Lett.\ B {\bf 225}, 363 (1989).
  
  
  
\bibitem{Derendinger:1991hq} 
  J.~P.~Derendinger, S.~Ferrara, C.~Kounnas and F.~Zwirner,
  Nucl.\ Phys.\ B {\bf 372}, 145 (1992).



\bibitem{Ibanez:1992hc} 
  L.~E.~Ibanez and D.~Lust,
  Nucl.\ Phys.\ B {\bf 382}, 305 (1992)
  [hep-th/9202046].


\bibitem{Araki:2007ss} 
  T.~Araki, K.~S.~Choi, T.~Kobayashi, J.~Kubo and H.~Ohki,
  Phys.\ Rev.\ D {\bf 76}, 066006 (2007)
  [arXiv:0705.3075 [hep-ph]].


\bibitem{Kobayashi:2016ovu} 
  T.~Kobayashi, S.~Nagamoto and S.~Uemura,
PTEP {\bf 2017}, no. 2, 023B02 (2017)
[arXiv:1608.06129 [hep-th]].



\bibitem{Ibanez:1991zv} 
  L.~E.~Ibanez, D.~Lust and G.~G.~Ross,
  Phys.\ Lett.\ B {\bf 272}, 251 (1991)
  [hep-th/9109053].


\bibitem{Kawabe:1994mj} 
  H.~Kawabe, T.~Kobayashi and N.~Ohtsubo,
  Nucl.\ Phys.\ B {\bf 434}, 210 (1995)
  [hep-ph/9405420].



\bibitem{Lauer:1989ax} 
  J.~Lauer, J.~Mas and H.~P.~Nilles,
  Phys.\ Lett.\ B {\bf 226}, 251 (1989);
%
  Nucl.\ Phys.\ B {\bf 351}, 353 (1991).
  

\bibitem{Lerche:1989cs} 
  W.~Lerche, D.~Lust and N.~P.~Warner,
  Phys.\ Lett.\ B {\bf 231}, 417 (1989).


\bibitem{Ferrara:1989qb} 
  S.~Ferrara, .D.~Lust and S.~Theisen,
  Phys.\ Lett.\ B {\bf 233}, 147 (1989).



\bibitem{Cremades:2004wa} 
  D.~Cremades, L.~E.~Ibanez and F.~Marchesano,
JHEP {\bf 0405}, 079 (2004)
[hep-th/0404229].



\bibitem{Kobayashi:2017dyu} 
  T.~Kobayashi and S.~Nagamoto,
  Phys.\ Rev.\ D {\bf 96}, no. 9, 096011 (2017)
  [arXiv:1709.09784 [hep-th]].
  
\bibitem{Kobayashi:2018rad} 
  T.~Kobayashi, S.~Nagamoto, S.~Takada, S.~Tamba and T.~H.~Tatsuishi,
  Phys.\ Rev.\ D {\bf 97}, no. 11, 116002 (2018)
  [arXiv:1804.06644 [hep-th]].



\bibitem{Kobayashi:2018bff} 
  T.~Kobayashi and S.~Tamba,
  Phys.\ Rev.\ D {\bf 99}, no. 4, 046001 (2019)
  [arXiv:1811.11384 [hep-th]].


\bibitem{Kobayashi:2006wq} 
  T.~Kobayashi, H.~P.~Nilles, F.~Ploger, S.~Raby and M.~Ratz,
  Nucl.\ Phys.\ B {\bf 768}, 135 (2007)
  [hep-ph/0611020].


\bibitem{Kobayashi:2004ya} 
  T.~Kobayashi, S.~Raby and R.~J.~Zhang,
  Nucl.\ Phys.\ B {\bf 704}, 3 (2005)
  [hep-ph/0409098].


\bibitem{Ko:2007dz} 
  P.~Ko, T.~Kobayashi, J.~h.~Park and S.~Raby,
  Phys.\ Rev.\ D {\bf 76}, 035005 (2007)
  Erratum: [Phys.\ Rev.\ D {\bf 76}, 059901 (2007)]
  [arXiv:0704.2807 [hep-ph]].


\bibitem{Beye:2014nxa} 
  F.~Beye, T.~Kobayashi and S.~Kuwakino,
  Phys.\ Lett.\ B {\bf 736}, 433 (2014)
  [arXiv:1406.4660 [hep-th]].


\bibitem{Olguin-Trejo:2018wpw} 
  Y.~Olguin-Trejo, R.~Perez-Martinez and S.~Ramos-Sanchez,
  Phys.\ Rev.\ D {\bf 98}, no. 10, 106020 (2018)
  [arXiv:1808.06622 [hep-th]].


\bibitem{Nilles:2018wex} 
  H.~P.~Nilles, M.~Ratz, A.~Trautner and P.~K.~S.~Vaudrevange,
  Phys.\ Lett.\ B {\bf 786}, 283 (2018)
  [arXiv:1808.07060 [hep-th]].




\bibitem{Abe:2009vi} 
  H.~Abe, K.~S.~Choi, T.~Kobayashi and H.~Ohki,
  Nucl.\ Phys.\ B {\bf 820}, 317 (2009)
  [arXiv:0904.2631 [hep-ph]].


\bibitem{Abe:2009uz} 
  H.~Abe, K.~S.~Choi, T.~Kobayashi and H.~Ohki,
  Phys.\ Rev.\ D {\bf 80}, 126006 (2009)
  [arXiv:0907.5274 [hep-th]];
%
  Phys.\ Rev.\ D {\bf 81}, 126003 (2010)
  [arXiv:1001.1788 [hep-th]].


\bibitem{BerasaluceGonzalez:2012vb} 
  M.~Berasaluce-Gonzalez, P.~G.~Camara, F.~Marchesano, D.~Regalado and A.~M.~Uranga,
  JHEP {\bf 1209}, 059 (2012)
  [arXiv:1206.2383 [hep-th]].


\bibitem{Marchesano:2013ega} 
  F.~Marchesano, D.~Regalado and L.~Vazquez-Mercado,
  JHEP {\bf 1309}, 028 (2013)
  [arXiv:1306.1284 [hep-th]].
  
  
\bibitem{Abe:2014nla} 
  H.~Abe, T.~Kobayashi, H.~Ohki, K.~Sumita and Y.~Tatsuta,
  JHEP {\bf 1406}, 017 (2014)
  [arXiv:1404.0137 [hep-th]].
  

\bibitem{Higaki:2005ie} 
  T.~Higaki, N.~Kitazawa, T.~Kobayashi and K.~j.~Takahashi,
  Phys.\ Rev.\ D {\bf 72}, 086003 (2005)
  [hep-th/0504019].
  
  

  





\bibitem{Baur:2019kwi} 
  A.~Baur, H.~P.~Nilles, A.~Trautner and P.~K.~S.~Vaudrevange,
  arXiv:1901.03251 [hep-th].







\bibitem{Altarelli:2010gt} 
  G.~Altarelli and F.~Feruglio,
  Rev.\ Mod.\ Phys.\  {\bf 82}, 2701 (2010)
  [arXiv:1002.0211 [hep-ph]].


\bibitem{Ishimori:2010au} 
  H.~Ishimori, T.~Kobayashi, H.~Ohki, Y.~Shimizu, H.~Okada and M.~Tanimoto,
  Prog.\ Theor.\ Phys.\ Suppl.\  {\bf 183}, 1 (2010)
  [arXiv:1003.3552 [hep-th]]; 
%
  Lect.\ Notes Phys.\  {\bf 858}, 1 (2012).


\bibitem{King:2013eh} 
  S.~F.~King and C.~Luhn,
  Rept.\ Prog.\ Phys.\  {\bf 76}, 056201 (2013)
  [arXiv:1301.1340 [hep-ph]].
  




\bibitem{deAdelhartToorop:2011re} 
  R.~de Adelhart Toorop, F.~Feruglio and C.~Hagedorn,
  Nucl.\ Phys.\ B {\bf 858}, 437 (2012)
  [arXiv:1112.1340 [hep-ph]].


\bibitem{Feruglio:2017spp} 
  F.~Feruglio,
  arXiv:1706.08749 [hep-ph].




\bibitem{Kobayashi:2018vbk} 
  T.~Kobayashi, K.~Tanaka and T.~H.~Tatsuishi,
  Phys.\ Rev.\ D {\bf 98}, no. 1, 016004 (2018)
  [arXiv:1803.10391 [hep-ph]].



\bibitem{Penedo:2018nmg} 
  J.~T.~Penedo and S.~T.~Petcov,
  Nucl.\ Phys.\ B {\bf 939}, 292 (2019)
  [arXiv:1806.11040 [hep-ph]].
  
  
\bibitem{Criado:2018thu} 
  J.~C.~Criado and F.~Feruglio,
  SciPost Phys.\  {\bf 5}, no. 5, 042 (2018)
  [arXiv:1807.01125 [hep-ph]].
  
  
\bibitem{Kobayashi:2018scp} 
  T.~Kobayashi, N.~Omoto, Y.~Shimizu, K.~Takagi, M.~Tanimoto and T.~H.~Tatsuishi,
  JHEP {\bf 1811}, 196 (2018)
  [arXiv:1808.03012 [hep-ph]].


\bibitem{Novichkov:2018ovf} 
  P.~P.~Novichkov, J.~T.~Penedo, S.~T.~Petcov and A.~V.~Titov,
  arXiv:1811.04933 [hep-ph];
%
%
  arXiv:1812.02158 [hep-ph].
  
\bibitem{Okada:2018yrn} 
  H.~Okada and M.~Tanimoto,
  Phys.\ Lett.\ B {\bf 791}, 54 (2019)
  [arXiv:1812.09677 [hep-ph]].


\bibitem{Kobayashi:2018wkl} 
  T.~Kobayashi, Y.~Shimizu, K.~Takagi, M.~Tanimoto, T.~H.~Tatsuishi and H.~Uchida,
  arXiv:1812.11072 [hep-ph].


\bibitem{Novichkov:2018yse} 
  P.~P.~Novichkov, S.~T.~Petcov and M.~Tanimoto,
  arXiv:1812.11289 [hep-ph].


\bibitem{Ding-King-Liu} 
  G.~J.~Ding, S.~F.~King and X.~G.~Liu,
  arXiv:1903.12588 [hep-ph].

\bibitem{Krauss:1988zc} 
  L.~M.~Krauss and F.~Wilczek,
  Phys.\ Rev.\ Lett.\  {\bf 62}, 1221 (1989).


\bibitem{Ibanez:1991hv} 
  L.~E.~Ibanez and G.~G.~Ross,
  Phys.\ Lett.\ B {\bf 260}, 291 (1991).

\bibitem{Banks:1991xj} 
  T.~Banks and M.~Dine,
  Phys.\ Rev.\ D {\bf 45}, 1424 (1992)
  [hep-th/9109045].

\bibitem{Araki:2008ek} 
  T.~Araki, T.~Kobayashi, J.~Kubo, S.~Ramos-Sanchez, M.~Ratz and P.~K.~S.~Vaudrevange,
  Nucl.\ Phys.\ B {\bf 805}, 124 (2008)
  [arXiv:0805.0207 [hep-th]].



\bibitem{Aldazabal:2000dg} 
  G.~Aldazabal, S.~Franco, L.~E.~Ibanez, R.~Rabadan and A.~M.~Uranga,
  J.\ Math.\ Phys.\  {\bf 42}, 3103 (2001)
  [hep-th/0011073].
  
  
  
  
  


\bibitem{Abe:2008fi} 
  H.~Abe, T.~Kobayashi and H.~Ohki,
  JHEP {\bf 0809}, 043 (2008)
  [arXiv:0806.4748 [hep-th]].
  







\end{thebibliography}
\end{document}